\documentclass[conference,a4paper]{IEEEtran}
\usepackage{}
\usepackage[latin1]{inputenc}
\usepackage{amssymb}
\usepackage{latexsym}
\usepackage{mathrsfs}
\usepackage{amsfonts}
\usepackage{amsbsy}
\usepackage{amsmath}
\usepackage{times}
\usepackage{epsfig,epsf,times,amssymb,amsmath}
\usepackage{color}
\usepackage{setspace}

\newcommand{\argmin}{\operatornamewithlimits{argmin}}

\IEEEoverridecommandlockouts
\title{Study of Buffer-Aided Space-Time Coding for Multiple-Antenna Cooperative Wireless Networks}

\author{\IEEEauthorblockN{Tong Peng and Rodrigo C. de Lamare}
\IEEEauthorblockA{CETUC/PUC-RIO, BRAZIL\\ Communications Research Group, Department of Electronics, University of York, York YO10 5DD, UK\\Email: tong.peng@cetuc.puc-rio.br; delamare@cetuc.puc-rio.br}
\thanks{This research is supported by the National Council for Scientific and Technological Development (CNPq) in Brazil.}}

\begin{document}
\maketitle

\begin{abstract}
In this work we propose an adaptive buffer-aided space-time coding scheme for cooperative wireless networks. A maximum likelihood receiver and adjustable code vectors are considered subject to a power constraint with an amplify-and-forward cooperation strategy. Each multiple-antenna relay is equipped with a buffer and is capable of storing the received symbols before forwarding them to the destination. We also present an adaptive relay selection and optimization algorithm, in which the instantaneous signal to noise ratio in each link is calculated and compared at the destination. An adjustable code vector obtained by a feedback channel at each relay is employed to form a space-time coded vector which achieves a higher coding gain than standard schemes. A stochastic gradient algorithm is developed to compute the parameters of the adjustable code vector with reduced computational complexity. Simulation results show that the proposed buffer-aided scheme and algorithm obtain performance gains over existing schemes.
\end{abstract}

\begin{IEEEkeywords}
Cooperative systems, buffer-aided relays, space-time codes, relay selection.
\end{IEEEkeywords}

\section{Introduction}

Cooperative multiple-input multiple-output (MIMO) systems
\cite{mmimo}-\cite{wence}, which employ relay nodes with multiple antennas
between the source node and the destination node as a distributed antenna
array, can obtain diversity gains by providing copies of the transmitted
signals to improve the reliability of wireless communication systems
\cite{J.N.Laneman2004,Clarke1}. In traditional cooperative systems,
amplify-and-forward (AF), decode-and-forward (DF) or compress-and-forward (CF)
\cite{J.N.Laneman2004} cooperation strategies are designed with the help of
multiple relay nodes. Relay selection algorithms such as those designed in
\cite{Clarke1,Clarke2} provide an efficient way to assist the communication
between the source node and the destination node.

Although the best relay node can be selected according to different optimization criteria, the traditional relay selection focuses on the best relay selection (BRS) scheme \cite{A.Bletsas} which selects the links with maximum instantaneous signal-to-noise ratio ($SNR$). Recently a new cooperative scheme with a source, a destination and multiple relays equipped with buffers has been introduced and analyzed in \cite{N.Zlatanov}-\cite{A.Ikhlef2}. The main idea is to select the best link during each time slot according to different criteria, such as maximum instantaneous $SNR$ and maximum throughput. In \cite{N.Zlatanov}, an introduction to buffered relaying networks is given, and a further analysis of the throughput and diversity gain is provided in \cite{N.Zlatanov2}. In \cite{N.Zlatanov3, N.Zlatanov4}, an adaptive link selection protocol with buffer-aided relays is proposed and an analysis of the network throughput and outage probability is developed. A max-link relay selection scheme focused on achieving full diversity gain which selects the strongest link in each time slot is proposed in \cite{I.Krikidis}. A max-max relay selection algorithm is proposed in \cite{A.Ikhlef2} and has been extended to mimic a full-duplex relaying scheme in \cite{A.Ikhlef} with the help of buffer-aided relays.

In this paper, we propose an adjustable buffer-aided space-time coding (STC)
schemes and an adaptive buffer-aided relaying optimization (ABARO) algorithm
for cooperative relaying systems with feedback. The proposed algorithm can be
divided into two parts, the first one is the relay selection part which chooses
the best link with the maximum instantaneous $SNR$ and checks if the state of
the best relay node is available to transmit or receive, and the second part
refers to the optimization part for the adjustable STC schemes employed at the
relay nodes. The proposed algorithm is based on the maximum-likelihood (ML)
criterion subject to constraints on the transmitted power at the relays for
different cooperative systems. Due to the use of STC schemes at each relay
node, an ML detector is employed at the destination node in order to achieve
full receive diversity. Suboptimal detectors \cite{delamare_mber}-\cite{mbdf}
can be also used at the destination node to reduce the detection complexity as
well as various parameter estimation techniques could be considered
\cite{S.Haykin}-\cite{barc} and precoding \cite{keke1}-\cite{rob_mbthp} and
decoding approaches. Moreover, stochastic gradient (SG) estimation methods
\cite{S.Haykin} are developed in order to compute the required parameters at a
reduced computational complexity. We study how the adjustable code vectors can
be employed at the buffer-aided relays combined with the relay selection
process and how to optimize the adjustable code vectors by employing an ML
criterion. The proposed relay selection and designs can be implemented with
different types of STC schemes in cooperative relaying systems with DF or AF
protocols. The main differences of this work as compared to
\cite{N.Zlatanov}-\cite{A.Ikhlef2} are the use of STC schemes to improve the
accuracy of the transmission and the multiple antennas used at each node to
ensure the full diversity order of the STC schemes.

The paper is organized as follows. Section II introduces a cooperative two-hop relaying systems with multiple buffer-aided relays applying the AF strategy and adjustable STC schemes. In Section III the encoding and decoding procedure of the adjustable STC schemes are introduced and in Section IV the proposed relay selection and coding vector optimization algorithms are derived. The results of the simulations are given in Section V. Section VI gives the conclusions of the work.

Notation: the italic, the bold lower-case and the bold upper-case letters denote scalars, vectors and matrices, respectively. The operator $\parallel{\boldsymbol X}\parallel_F=\sqrt{{\rm Tr}({\boldsymbol X}^H\cdot{\boldsymbol X})}=\sqrt{{\rm Tr}({\boldsymbol X}\cdot{\boldsymbol X}^\emph{H})}$ is the Frobenius norm. ${\rm Tr}(\cdot)$ stands for the trace of a matrix, and the $N \times N$ identity matrix is written as ${\boldsymbol I}_N$.

\section{Cooperative System Model}

\begin{figure}
\begin{center}
\def\epsfsize#1#2{0.825\columnwidth}
\epsfbox{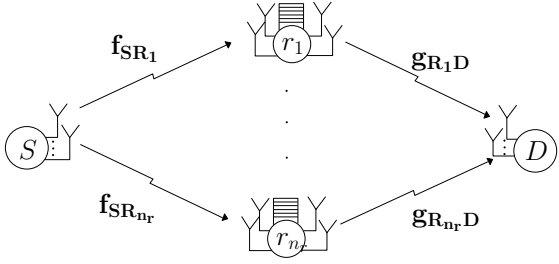} \caption{Cooperative System Model with $n_r$ Relay
Nodes}\label{f1} \vspace{-1em}
\end{center}
\end{figure}

We consider a two-hop cooperative communication system, which consists of one
source node, one destination node and $n_r$ relay nodes (Relay 1, Relay 2, ...,
Relay $n_r$) as shown in Fig. 1. All the nodes employ $N$ antennas and can
either transmit or receive at one time. Each relay node contains a buffer and
can store the received symbols if the buffer is not full. The key advantage of
employing relays with buffers is to increase the reliability of the
transmission by means of deciding the best time to forward data
\cite{N.Zlatanov}-\cite{A.Ikhlef2}. The two main challenges of using
buffer-aided relays are how to obtain accurate instantaneous CSI and how to
deal with the delay. The calculation of the instantaneous $SNR_{ins}$ and
comparisons are required before every transmission so that the key element of
choosing the best relay node or the relay sets is the accuracy of the CSI in
each link. The delay caused by the best relay selection strategy is another
problem to some types of information such as real-time transmission of video
and speech. However, the authors in \cite{N.Zlatanov} have observed an
improvement in performance due to the introduction of extra degrees of freedom
by using buffer-aided relaying systems as compared to the traditional
cooperative systems. Therefore, it is suggested that the applications of
buffer-aided relays could be used in cellular and sensor networks
\cite{N.Zlatanov}. In this work, we consider only one user at the source node
in our system that operates in a spatial multiplexing configuration, and we
assume that perfect CSI, as well as the states of each relay, are available at
the destination node. An appropriate signalling that provides global CSI at the
destination node can ensure this assumption \cite{I.Krikidis}. The destination
node will send out the relay selection information together with the code
vector via a feedback channel which is assumed error-free. The channel
considered in this paper is assumed to be static over some blocks,
non-selective Rayleigh block fading with additive white Gaussian noise (AWGN),
and the channel coefficients remain constant during one time slot and change
from one time slot to another. We assume the feedback required in the proposed
ABARO algorithm is short enough to be ignored as compared to the symbol
transmission period. The same assumption is made in
\cite{N.Zlatanov}-\cite{A.Ikhlef2}. If the cooperative system operates in
time-division duplex (TDD) mode then the feedback information can be obtained
at the transmitter subject to hardware differences and impairments.

Let ${\boldsymbol s}$ denote the modulated data symbol packet with $J$ symbols
and covariance matrix $E\big[{\boldsymbol s}{\boldsymbol s}^\emph{H}\big] =
\sigma_{s}^{2}{\boldsymbol I}_N$, where $\sigma_s^2$ denotes the signal power.
Assuming a channel that is static over one packet, the destination node
calculates the instantaneous $SNR$, $SNR_{ins}$, and then the optimal relay
node is selected by choosing the highest $SNR$ of the link. The information of
the optimal relay node selection is sent back to the relays via an error-free
feedback channel. It is assumed that the time for calculation of the $SNR$ and
selection feedback is short enough to be ignored compared to that for
transmission or reception. Then in the first time slot, the source node sends
the first $N$ modulated information symbols in ${\boldsymbol s}$ to the optimal
relay node. The expression of the received data in the first time slot is given
by
\begin{equation}\label{2.1}
\begin{aligned}
    \boldsymbol{r}_{SR_k}[i] &= \sqrt{\frac{P_S}{N}}\boldsymbol{F}_{SR_k}[i]\boldsymbol{s}[i] + \boldsymbol{n}_{SR_k}[i],\\ &k=1,2,...,n_r, ~i = 1,2,...,\frac{J}{N},
\end{aligned}
\end{equation}
where ${\boldsymbol F}_{SR_k}$ denotes the CSI between the source node and the $k$th relay, and $n_{SR_k}$ stands for the AWGN generated at the $k$th relay with variance $\sigma^2_r$. The time slot index is indicated by $i$. After reception, the destination node calculates the instantaneous $SNR$ of the \emph{SR} link and the \emph{RD} link, where the \emph{SR} link stands for the link between the source node and the relay node and the \emph{RD} links stands for the link between the relay node and the destination node. In order to select the best relay node for the second time slot and the selection information will be sent back to the selected relay node so that the relay will be ready for forwarding the information in the buffer to the destination node or for receiving data sent from the source node. We assume the best link is the $k$th \emph{RD} link and the buffer is not empty, and due to the AF strategy the received symbols at the best relay will be forwarded to the destination node without detection at the relay node. The $RD$ link transmission is expressed as follows:
\begin{equation}\label{2.2}
\begin{aligned}
    \boldsymbol{r}_{R_kD}[i] &= \sqrt{\frac{P_R}{N}}\boldsymbol{G}_{R_kD}[i]\boldsymbol{r}_{SR_k}[i] + \boldsymbol{n}_{R_kD}[i], \\ &k=1,2,...,n_r, ~i = 1,2,...,\frac{J}{N},
\end{aligned}
\end{equation}
where $\boldsymbol{G}_{R_kD}[i]$ denotes the CSI between the $k$th relay and the destination node, and $\boldsymbol{n}_{R_kD}$ stands for the AWGN vector generated at the destination node with variance $\sigma^2_d$.

\section{Adjustable Space-Time Coding Scheme}

As mentioned in the previous section, different STC schemes can be employed at the relay nodes to achieve an improvement in BER performance. In this work, randomized space-time coding (RSTC) schemes in \cite{B.Sirkeci-Mergen} are considered and the adaptive optimization algorithms in \cite{TARMO} are also employed to achieve an improvement in BER performance. At each relay node, an adjustable code vector is randomly generated before the forwarding procedure in order to adapt different STC schemes into single-antenna relays. For example, when the buffer size at the relays is equal to $2$ this indicates simple STC schemes, such as $2 \times 2$ Alamouti space-time block code (STBC) can be implemented at the relays. By multiplying a $2 \times 1$ code vector, the original $2 \times 2$ Alamouti scheme changes to a $2 \times 1$ adjustable code vector and can be forwarded to the destination node within $2$ time slots. The use of an adjustable code vector provides lower BER performance and higher diversity gain \cite{B.Sirkeci-Mergen}\cite{TARMO}. The received data matrix is described by
\begin{equation}\label{3.1}
\begin{aligned}
    \boldsymbol{R}_{R_kD}[i] &= \sqrt{\frac{P_R}{N}}\boldsymbol{G}_{R_kD}[i]{\boldsymbol C}_{rand}[i] + \boldsymbol{N}_{R_kD}[i], \\ &~k=1,2,...,n_r, ~i = 1,2,...,\frac{J}{N},
\end{aligned}
\end{equation}
where $\boldsymbol{R}_{R_kD}[i]$ denotes the $N \times T$ received symbol matrix and $T$ is the transmission time slot of the STC scheme. The $N \times T$ adjustable STC matrix is denoted by ${\boldsymbol C}[i]$, and $\boldsymbol{G}_{R_kD}$ denotes the CSI matrix between the $k$th relay and the destination node. $\boldsymbol{N}_{R_kD}$ stands for the AWGN matrix generated at the destination node with variance $\sigma^2_d$. In this paper, we assume that the channel does not change during the transmission of one adjustable STC scheme. Since the STC is employed at relay nodes, the received data ${\boldsymbol R}_{R_kD}[i]$ in ({\ref{3.1}}) can be rewritten as
\begin{equation}\label{3.2}
\begin{aligned}
    {\boldsymbol r}_{R_kD}[i] =& \sqrt{\frac{P_R P_S}{N}}{\boldsymbol V}_{eq}[i]{\boldsymbol H}[i]{\boldsymbol s}[i] + \sqrt{\frac{P_R}{N}}{\boldsymbol V}_{eq}[i]{\boldsymbol G}_{R_kD}[i]\\&{\boldsymbol n}_{SR_k}[i] + {\boldsymbol n}_{R_kD}[i]\\ =& \sqrt{\frac{P_R}{N}}{\boldsymbol V}_{eq}[i]{\boldsymbol H}[i]{\boldsymbol s}[i] + {\boldsymbol n}[i],
\end{aligned}
\end{equation}
where ${\boldsymbol V}_{eq}$ denotes the $TN \times TN$ block diagonal equivalent adjustable code matrix, and ${\boldsymbol H}[i]$ stands for the equivalent channel matrix which is the combination of ${\boldsymbol F}_{SR_k}[i]$ and ${\boldsymbol G}_{R_kD}[i]$. The vector ${\boldsymbol n}[i]$ contains the equivalent received noise vector at the destination node, which can be modeled as AWGN with zero mean and covariance matrix $(\sigma^{2}_d+\|{\boldsymbol V}_{eq}[i]{\boldsymbol G}_{R_kD}[i]\|^2_F\sigma^{2}_r){\boldsymbol I}_{NT}$.

The number of antennas is $N$ and the packet size is $J$, as a result, during the transmission, the packet is divided into $i=J/N$ groups if the size of packet $J$ is larger than the number of antennas $N$ and we assume $i$ is an integer. For example, the packet size $J$ is much larger than the number of antennas $N$, and $N=2$ to implement the $2 \times 2$ Alamouti STBC scheme at the relays.

In the first hop, we divide the $J \times 1$ symbol vector ${\boldsymbol s}[i]$ into $2$ groups of $J/2 \times 1$ sub-vectors and then transmit them to the relay node from the $2$ transmit antennas. At the relays, we first divide ${\boldsymbol r}_{SR_k}$ into $2$ groups and then multiplied by a $2 \times 2$ diagonal adjustable code matrix. After that the original $2 \times 2$ orthogonal Alamouti STBC scheme changes to the following code vector:
\begin{equation}\label{3.3}
\begin{aligned}
    {\boldsymbol C}_{rand} &= {\boldsymbol V}{\boldsymbol C} = \left[\begin{array}{cc} v_1 & 0 \\ 0 & v_2\end{array}\right]\left[\begin{array}{cc} r_{SR_k}1 & -r^*_{SR_k}2 \\ r_{SR_k}2 & ~r^*_{SR_k}1 \end{array} \right]\\ & = \left[\begin{array}{cc}v_1r_{SR_k}1 & -v_1r^*_{SR_k}2 \\ v_2r_{SR_k}2 & v_2r^*_{SR_k}1\end{array}\right],
\end{aligned}
\end{equation}
where $r_{SR_k}1$ and $r_{SR_k}2$ are the first symbols in the separate groups, and the $2 \times 2$ matrix ${\boldsymbol V}$ denotes the randomized matrix whose elements in the diagonal are generated randomly according to different criteria described in \cite{B.Sirkeci-Mergen}.

\section{Adaptive Buffer-Aided STC And Relaying Optimization Algorithm}

In this section, the proposed ABARO algorithm is derived in detail. Before each transmission, the instantaneous $SNR$ of the $SR$ and $RD$ links is calculated at the destination, which is given by
\begin{equation}\label{4.1}
    SNR_{SR_k} = \frac{\sqrt{\|{\boldsymbol F}_{SR_k}\|^2_F}}{\sigma^2_r}, ~SNR_{R_kD} = \frac{\sqrt{\|{\boldsymbol V}_{eq}{\boldsymbol G}_{R_kD}\|^2_F}}{\sigma^2_d},
\end{equation}
and the best link is chosen according to
\begin{equation}\label{4.2}
    SNR_{\rm opt} = \arg\max_{k} SNR_{{\rm ins}_k}, k=1,2,~...~,n_r.
\end{equation}
Once the best relay is determined, the transmission is described as (\ref{2.1}). After the reception, the destination node calculates the instantaneous $SNR$ in the $SR$ links and $RD$ links, respectively, and chooses the best link for the next time slot. At the relay node, RSTC schemes are employed in order to enhance the transmission. Therefore, the calculation of the instantaneous $SNR$ for the links between the relay nodes and the destination should contain the adjustable code vector in \cite{TARMO}, as described by
\begin{equation}\label{4.3.1}
    SNR_{R_kD} = \frac{\sqrt{\|{\boldsymbol V}_{eq}{\boldsymbol G}_{R_kD}\|^2_F}}{\sigma^2_d}.
\end{equation}
The relay states are known at the destination node so that if the $k$th $RS$ link is chosen but the buffer at the $k$th relay node is empty, the source node will skip this node and check the state of the buffer which has the second best link. In this case the optimal relay may not be chosen at each time slot but the delay period will be decreased. The process repeats until the last information symbol is received at the destination node. After the detection at the destination node, the adjustable code matrix ${\boldsymbol V}$ will be optimized and updated. The constrained ML optimization problem can be written as
\begin{equation}\label{4.3}
\begin{aligned}
     & \left[\hat {\boldsymbol s}[i], \hat {\boldsymbol V}_{eq}[i]\right]  = \argmin_{{\boldsymbol s}[i],{\boldsymbol V}_{eq}[i]} \|{\boldsymbol r}_{R_kD}[i] - \sqrt{\frac{P_R P_S}{N}}{\boldsymbol V}_{eq}[i]{\boldsymbol H}[i]\hat{\boldsymbol s}_{SR_k}[i]\|^2, \\
     & ~~~~~~~~~~~~~~~~~~~~~~s.t.~ {\rm Tr}({\boldsymbol V}_{eq}[i]{\boldsymbol V}^\emph{H}_{eq}[i])\leq {P_C},
\end{aligned}
\end{equation}
where $\hat{\boldsymbol s}_{SR_k}[i]$ denotes the detected symbol vector forwarded from the $k$th relay node. According to the property of the adjustable code matrix, the computation of $\hat{\boldsymbol s}[i]$ is the same as the decoding procedure of the original STC schemes. In order to obtain the optimal coding matrix ${\boldsymbol V}[i]$, the cost function in (\ref{4.3}) should be minimized with respect to the equivalent code matrix ${\boldsymbol V}_{eq}[i]$ subject to a constraint on the transmitted power. The Lagrangian expression of the optimization problem in (\ref{4.3}) is given by
\begin{equation}\label{4.4}
\begin{aligned}
    \mathcal {L} =& \|{\boldsymbol r}_{R_kD}[i] - \sqrt{\frac{P_R P_S}{N}}{\boldsymbol V}_{eq}[i]{\boldsymbol H}[i]\hat{\boldsymbol s}_{SR_k}[i]\|^2\\ & + \lambda(Tr({\bf V}_{\rm eq}[i] {\bf V}_{\rm eq}^{H}[i]) - P_v).
\end{aligned}
\end{equation}
A stochastic gradient algorithm can be used to solve the optimization algorithm in (\ref{4.3}) or equivalently to minimize (\ref{4.4}) with lower computational complexity as compared to least-squares algorithms which require the inversion of matrices. A normalization procedure to enforce the power constraint. By taking the instantaneous gradient of $\mathcal {L}$, discarding the power constraint and equating it to zero, we can obtain
\begin{equation}\label{4.5}
\begin{aligned}
    \nabla\mathcal {L} = & -\sqrt{\frac{P_R P_S}{N}}\times\\&({\boldsymbol r}_{R_kD}[i] - \sqrt{\frac{P_R P_S}{N}}{\boldsymbol V}_{eq}[i]{\boldsymbol H}[i]\hat{\boldsymbol s}_{SR_k}[i])\hat{\boldsymbol s}^\emph{H}_{SR_k}[i]{\boldsymbol H}^\emph{H}[i],
\end{aligned}
\end{equation}
and the ABARO algorithm for the proposed scheme can be expressed as follows
\begin{equation}\label{4.6}
\begin{aligned}
    {\boldsymbol V}_{eq}[i+1] = & {\boldsymbol V}_{eq}[i] - \mu\sqrt{\frac{P_R P_S}{N}}({\boldsymbol r}_{R_kD}[i] - \\ & \sqrt{P_R}{\boldsymbol V}_{eq}[i]{\boldsymbol H}[i]\hat{\boldsymbol s}_{SR_k}[i])\hat{\boldsymbol s}^\emph{H}_{SR_k}[i]{\boldsymbol H}^\emph{H}[i],
\end{aligned}
\end{equation}
where $\mu$ is the step size. The normalization of the code vector ${\boldsymbol C}[i]$ instead of considering the power constraint in (\ref{4.5}) is given by
\begin{equation}\label{3.6}
    {\boldsymbol V}[i+1] = {\boldsymbol V}[i+1]\frac{P_V}{\sqrt{\|{\boldsymbol V}[i+1]\|_F^2}}.
\end{equation}

The summary of the ABARO algorithm is shown in Table I.

\begin{table*}
  \centering
  \caption{Summary of the Adaptive Buffer-Aided Relaying Optimization Algorithm}\label{t1}
  \begin{tabular}{|l|}
  \hline
  \bfseries{Initialization:}\\
  \qquad Empty the buffer at the relays,\\
  \bfseries{for $i=1,2,...,NK$} \\

  \qquad \bfseries{if $i=1$} \\
  \qquad\qquad compute: $SNR_{SR_k} = \frac{\sqrt{\|\boldsymbol{F}_{SR_k}[i]\|_F^2}}{\sigma^2_n}, ~k=1,2,...,n_r$\\
  \qquad\qquad compare: $SNR_{opt} = \arg\max \{SNR_{SR_k}\}$, ~$k=1,2, ... ,n_r$,\\
  \qquad\qquad $\boldsymbol{r}_{SR_k}[i] = \sqrt{\frac{P_S}{N}}\boldsymbol{F}_{SR_k}[i]\boldsymbol{s}[i] + \boldsymbol{n}_{SR_k}[i]$,\\

  \qquad \bfseries{else}\\
  \qquad\qquad compute: $SNR_{sr_k} = \frac{\sqrt{\|\boldsymbol{F}_{SR_k}[i]\|_F^2}}{\sigma^2_n}, ~k=1,2,...,n_r$\\
  \qquad\qquad\qquad\qquad $SNR_{R_kD} = \frac{\sqrt{\|{\boldsymbol V}_{eq}[i]{\boldsymbol G}_{R_kD}[i]\|^2_F}}{\sigma^2_d}, ~k=1,2,...,n_r$,\\
  \qquad\qquad compare: $SNR_{opt} = \arg\max \{SNR_{SR_k},SNR_{R_kD}\}$, ~$k=1,2, ... ,n_r$,\\
  \qquad\qquad\qquad \bfseries{if} $SNR_{max} = SNR_{SR_k} ~\& ~Relay_k \rm{ ~is ~not ~full}$\\
  \qquad\qquad\qquad\qquad $\boldsymbol{r}_{SR_k}[i] = \sqrt{\frac{P_S}{N}}\boldsymbol{F}_{SR_k}[i]\boldsymbol{s}[i] + \boldsymbol{n}_{SR_k}[i]$,\\
  \qquad\qquad\qquad \bfseries{elseif} $SNR_{max} = \arg\max SNR_{r_kd} ~\& ~Relay_k \rm{ ~is ~not ~empty}$\\
  \qquad\qquad\qquad\qquad ${\boldsymbol r}_{R_kD}[i] = \sqrt{\frac{P_R}{N}}{\boldsymbol V}_{eq}[i]{\boldsymbol H}[i]{\boldsymbol s}[i] + {\boldsymbol n}[i]$,\\
  \qquad\qquad\qquad\qquad ML detection: \\
  \qquad\qquad\qquad\qquad\qquad $\hat{\boldsymbol s}[i] = \arg\min_{{\boldsymbol s}[i]} \|{\boldsymbol r}_{R_kD}[i] - \sqrt{P_R}{\boldsymbol V}_{eq}[i]{\boldsymbol H}[i]\hat{\boldsymbol s}_{sr_k}[i]\|^2$,\\
  \qquad\qquad\qquad\qquad Adjustable Matrix Optimization: \\
  \qquad\qquad\qquad\qquad\qquad ${\boldsymbol V}_{eq}[i+1] = {\boldsymbol V}_{eq}[i] - \mu\sqrt{\frac{P_R P_S}{N}}({\boldsymbol r}_{R_kD}[i] - \sqrt{P_R}{\boldsymbol V}_{eq}[i]{\boldsymbol H}[i]\hat{\boldsymbol s}_{SR_k}[i])\hat{\boldsymbol s}^\emph{H}_{SR_k}[i]{\boldsymbol H}^\emph{H}[i]$,\\
  \qquad\qquad\qquad\qquad Normalization: \\
  \qquad\qquad\qquad\qquad\qquad ${\boldsymbol V}[i+1] = {\boldsymbol V}[i+1]\frac{P_V}{\sqrt{\|{\boldsymbol V}[i+1]\|_F^2}}$,\\
  \qquad\qquad\qquad \bfseries{elseif} $SNR_{sr_k} \rm{~is ~max} ~\& ~Relay_k \rm{ ~is ~full}$\\
  \qquad\qquad\qquad\qquad skip this Relay,\\
  \qquad\qquad\qquad \bfseries{elseif} $SNR_{r_kd} \rm{~is ~max} ~\& ~Relay_k \rm{ ~is ~empty}$\\
  \qquad\qquad\qquad\qquad skip this Relay,\\
  \qquad\qquad\qquad\qquad\qquad ...repeat...\\
  \qquad\qquad\qquad \bfseries{end}\\
  \qquad \bfseries{end}\\
  \bfseries{end}\\
  \bfseries{for} $d = 1,2,...,n_r$\\
  \qquad \bfseries{if} $Relay_{d} ~is ~not ~empty$,\\
  \qquad\qquad transmission: $Relay_{d} \rightarrow Destination$,\\
  \qquad \bfseries{end}\\
  \bfseries{end}\\

\hline
\end{tabular}
\end{table*}

\section{Simulation}

The simulation results are provided in this section to assess the proposed scheme and ABARO algorithm. In this work, we consider a multiple-antenna cooperative system that employs the AF protocol with the randomized Alamouti (R-Alamouti) in \cite{B.Sirkeci-Mergen}. The BPSK modulation is employed and each link between the nodes is characterized by static block fading with AWGN. It is possible to employ different STC schemes with a simple modification and to incorporate the proposed algorithm. We employ $n_r=1,2$ relay nodes and $N=2$ antennas at each node, and we set the symbol power $\sigma^2_s$ to 1. The packet size is $J=100$ with each packet containing $100$ symbols. The buffer size at the relays is equal to $J$.

The proposed ABARO algorithm with the Alamouti scheme and an ML receiver is evaluated with a two-relay system in Fig. 2. It is shown in the figure that the buffer-aided relay selection systems achieve $3$dB to $5$dB gains compared to the traditional relaying systems. When the BSR algorithm is considered at the relay node, an improvement of diversity order is shown in Fig. 2 which leads to dramatically improved BER performance. According to the simulation results in Fig. 2, a $1$dB gain can be achieved by using the RSTC scheme at the relays as compared to the network using the standard STC scheme at the relay node. When the proposed ABARO algorithm is employed at the relays, a $2$dB saving for the same BER performance as compared to the standard STC encoded system can be observed. The diversity order of using the proposed ABARO algorithm is the same as that of using the RSTC scheme at the relay node.

\begin{figure}
\begin{center}
\def\epsfsize#1#2{0.82\columnwidth}
\epsfbox{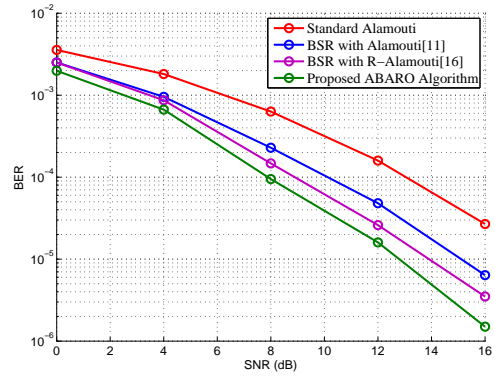} \caption{BER Performance vs. $SNR$ for Buffer-Aided Relaying
System}\label{f2}
\end{center}
\end{figure}

\section{Conclusion}

We have proposed a buffer-aided space-time coding scheme and the ABARO
algorithm for cooperative systems with feedback using an ML receiver at the
destination node to achieve a better BER performance. Simulation results have
illustrated the advantage of using the STC schemes in the buffer-aided
cooperative systems compared to the BRS algorithms. In addition, the proposed
ABARO algorithm can achieve a better performance in terms of lower bit error
rate at the destination node compare to the STC-ed systems. The ABARO algorithm
can be used with different STC schemes and can also be extended to cooperative
systems with any number of antennas.

\bibliographystyle{IEEEbib}
\bibliography{strings,refs}

\bibliographystyle{IEEEtran}

\end{document}